\documentclass[journal,comsoc]{IEEEtran}
\usepackage[T1]{fontenc}% optional T1 font encoding
\usepackage{amsmath}
\usepackage[cmintegrals]{newtxmath}
\usepackage{url}
\usepackage{soul,color,graphicx,float,epstopdf}
\usepackage{tablefootnote,textcomp,gensymb}
\usepackage{eurosym,booktabs,multirow}
\usepackage[caption=false]{subfig}
\usepackage{amsfonts} 
\usepackage{algorithmic}
\usepackage{array}
\usepackage{stfloats}
\usepackage{float}
\usepackage{mathtools}
\usepackage{graphicx}
\usepackage{pdfpages}
\usepackage{comment}
\usepackage{balance}
\usepackage{xcolor}
\usepackage{comment,enumitem}
\usepackage[normalem]{ulem}
\usepackage[implicit=false]{hyperref}

\definecolor{Orange}{rgb}{1,0.5,0}
\newcommand{\todo}[1]{\textsf{\textbf{\textcolor{Orange}{[#1]}}}}

\begin{document}
\title{Adapting MLOps for Diverse In-Network Intelligence in 6G Era: Challenges and Solutions}

\author{
% Authors
Peizheng~Li,~\IEEEmembership{}
         % Haiyuan~Li,~\IEEEmembership{}
         Ioannis Mavromatis,~\IEEEmembership{}
         Tim Farnham,~\IEEEmembership{}
         Adnan~Aijaz,~\IEEEmembership{}
         Aftab~Khan~\IEEEmembership{}
 \thanks{P. Li, T. Farnham, A. Aijaz, and A. Khan are with  Toshiba Europe Ltd., U. K.  (e-mail: peizheng.li@toshiba-bril.com).
 
 Ioannis Mavromatis is with Digital Catapult, NW1 2RA, London, U.K. This work was done while he was at Toshiba Europe Ltd., U.K.}
}
\maketitle

\begin{abstract}
Seamless integration of artificial intelligence (AI) and machine learning (ML) techniques with wireless systems is a crucial step for 6G \emph{AInization}. However, such integration faces challenges in terms of model functionality and lifecycle management. ML operations (MLOps) offer a systematic approach to tackle these challenges. 
Existing approaches toward implementing MLOps in a centralized platform often overlook the challenges posed by diverse learning paradigms and network heterogeneity.
This article provides a new approach to MLOps targeting the intricacies of future wireless networks. 
Considering unique aspects of the future radio access network (RAN), we formulate three operational pipelines, namely reinforcement learning operations (RLOps), federated learning operations (FedOps), and generative AI operations (GenOps). 
These pipelines form the foundation for seamlessly integrating various learning/inference capabilities into networks.
We outline the specific challenges and proposed solutions for each operation, facilitating large-scale deployment of AI-Native 6G networks.
\end{abstract}


\section{Introduction}
\label{sec:introduction}
Artificial intelligence (AI) and machine learning (ML) techniques have been shaping various industrial segments. Wireless networks are also proactively embracing AI/ML techniques from the perspectives of hardware, software, protocol, and standardization. For next-generation 6G networks, it is envisaged that AI/ML-based designs will facilitate network protocols, improve operational efficiency, and embed native intelligence and sustainability~\cite{letaief2019roadmap}. To corroborate, the ITU-R recently agreed on recommendations for the “IMT-2030 Framework''~\cite{ituimt2030}, wherein  AI and communication are regarded as the pillars for 6G. The framework defines the development, standardization, and deployment of 6G networks, setting the stage for transforming conventional technological landscape. 

On the standardization front, the ETSI operational coordination group on AI (OCG AI) is actively standardizing AI across various sectors and architectural models. ETSI zero-touch service management (ZSM) leverages AI as a key enabler for 5G/6G automation~\cite{rezazadeh2024sliceops}. In 3GPP, nearly all working groups (WGs) are now engaged in AI/ML-related activities. 
The objective is to manage the entire lifecycle of in-network AI/ML models, from training and emulation to deployment and inference.
3GPP WG SA2 has studied the distribution, transfer, and training of AI/ML models for various applications~\cite{3gpp23.700}, while SA5 has documented AI/ML management specifications and described a generic operational workflow~\cite{3gpp28.105,3gpp_website}. Considerable efforts have been dedicated to establishing a general AI/ML framework for the air interface.
The recent open radio access network (Open RAN) initiatives consider AI/ML capabilities crucial for innovation~\cite{polese2023understanding}. The O-RAN architecture further integrates AI/ML into the network. The RAN intelligent controllers (RICs) provide the base for AI/ML model deployment and actuation; and different AI/ML models can be deployed depending on the reaction time of the target task.

On the technological front, various new research activities are relying on AI-native design of future wireless systems. A prominent example is semantic and goal-oriented communications which leverages AI/ML for so-called \emph{beyond-Shannon} capabilities~\cite{strinati20216g}.
Meanwhile, recent advancements in generative AI for contextual understanding and content generation trigger enthusiasm for its integration with wireless networks. By doing so, wireless networks can be equipped with understanding and reasoning capabilities, improving their efficiency and robustness. 

The generalization ability of AI/ML remains a long-standing issue. The AI/ML models can merely work well for a specific task with constant training and inference features. This poses one critical question for the in-network operational AI/ML, i.e., ``how to ensure the effectiveness of the AI/ML models over complex and diversified wireless environments?'' Certainly, adapting the AI/ML models to different tasks using learning techniques such as domain adaption, transfer learning, or incremental learning is a proven solution. However, for this in-network deployed AI/ML application, when and how to trigger this adaptation is a more important question. A wireless network needs methods to measure and monitor the performance of AI/ML models and update these models accordingly. This is also termed the AI/ML model \emph{lifecycle management} (LCM), aiming to ensure that AI/ML takes effect in the wireless infrastructure fully controllable in their lifetime.

ML operations (MLOps) is a solution for AI/ML LCM which is based on operational principles from data engineering, AI/ML model, and continuous integration and continuous delivery/deployment (CI/CD) pipeline. It works well for AI/ML applications in a streamlined engineering environment with relatively simple tasks like supervised learning tasks. However, the circumstances are getting more complicated when applying other learning paradigms to realistic wireless communication systems. Each typical learning paradigm has its own peculiarities and challenges in implementation, which would be echoed or amplified with wireless system features. Current network virtualization deployments rely upon cloud-native tools and frameworks for developing and deploying applications in wireless networks. Short-living containerized applications can be automatically deployed and managed using orchestration tools such as Kubernetes. The specific data processing, model evaluation and updating requirements of different AI/ML models are usually overlooked in such deployments, hindering the service-assurance capability of in-network AI.

This article focuses on crafting in-network AI/ML model MLOps pipelines corresponding to different learning paradigms. We believe this could serve as the organic substrate for forming AI-Native capability. The key contributions are summarized as follows.
\begin{itemize}
    \item We systematically analyze the system and pipeline requirements of integrating reinforcement learning (RL), federated learning (FL), and generative AI (GenAI) into the wireless network, from the perspective of LCM by identifying the uniqueness of in-network functioning of AI/ML models.
    \item We raise customized Ops techniques to adapt the above learning paradigms: RLOps, FedOps, and GenOps.
    \item The features and best practices of in-wireless systems RLOps, FedOps, and GenOps are presented according to the engineering experiences. 
    \item We linked these Ops frameworks with specific vertical applications: RLOps is discussed for intelligent management and orchestration in Open RAN; FedOps is designed to cater to the requirements of large-scale IoT devices and edge intelligence; GenOps refers to the process of integrating generative models into the network broadly.    
\end{itemize}

\section{Preliminaries}
\label{sec:preliminary}
\subsection{MLOps}
\label{subsec:mlops}
\begin{figure}[t]
    \centering
    \includegraphics[width=0.8\linewidth]{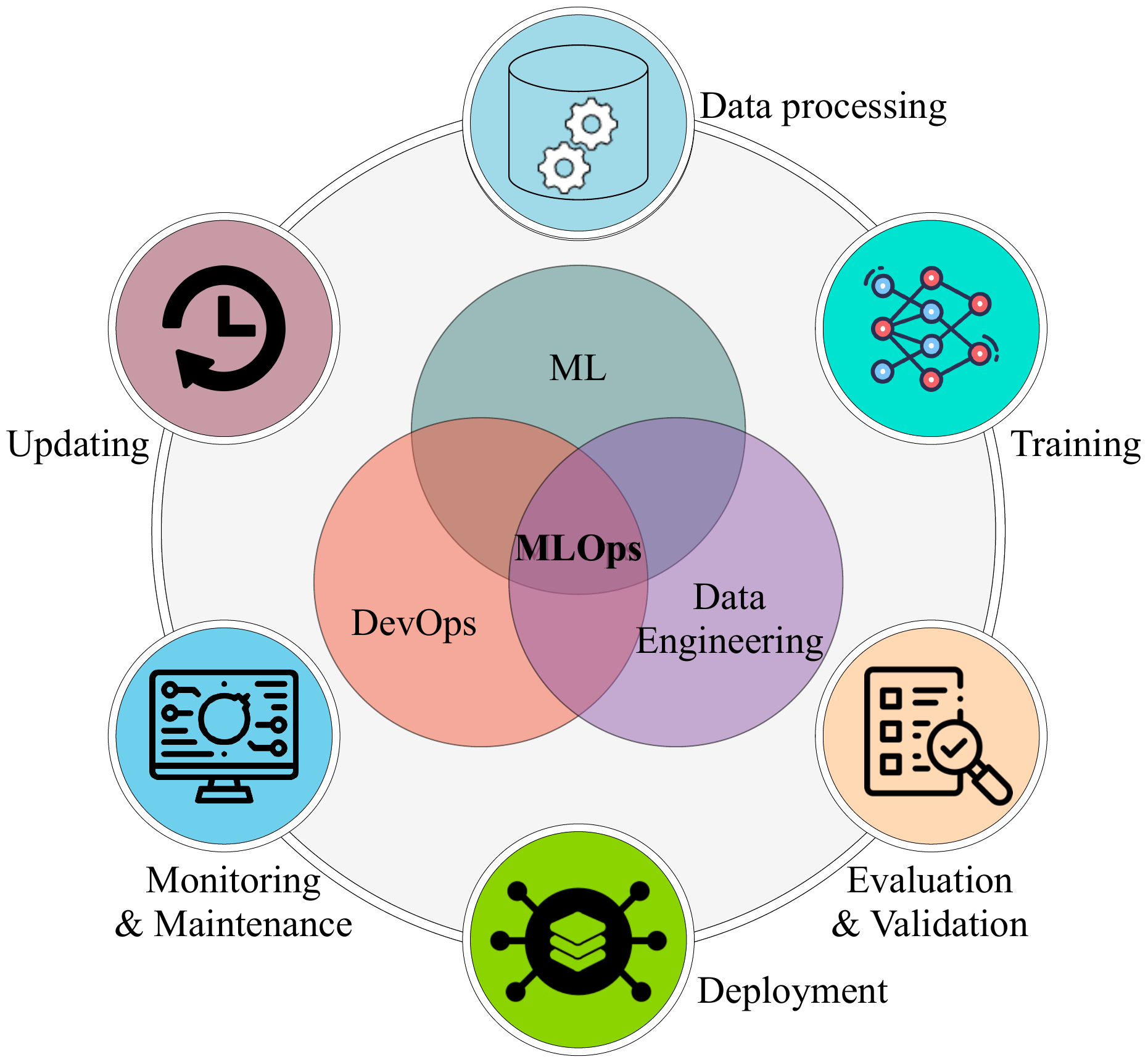}
    \vspace{-0.2cm}
    \caption{The key components comprising MLOps.}
    \vspace{-0.5cm}
    \label{fig:MLOps}
\end{figure}
\subsubsection{DevOps}
It is a foundational element within the MLOps framework, merging software development (Dev) and IT operations (Ops) to enhance the efficiency, quality, and speed of software development and delivery. DevOps methodology is widely adopted across the software development and IT industry. As a set of practices and tools, DevOps streamlines and automates the collaboration between software development and IT operations, improving and shortening the systems development life cycle. It emphasizes collaboration, communication, and process automation to facilitate faster and more reliable software development, testing, deployment, and maintenance. DevOps aims to dismantle barriers between development and operations, fostering an integrated and continuous approach to software development and deployment. 

\subsubsection{Data engineering}
Data engineering is a specialized domain within data science and analytics focused on the processes of data collection and processing. It involves designing, constructing, and managing the infrastructure and systems needed for gathering, storing, and organizing data for analysis, reporting, and other data-driven activities. Data engineers work with diverse data sources and technologies to ensure data is accessible and reliable. In the context of AI/ML, data engineering plays a crucial role in two key aspects:
\begin{enumerate}
    \item \textit{High-Quality Training Data Generation:} Data engineering supports the collection, filtering, and annotation processes to create high-quality training datasets.
    \item \textit{Continuous Data Capturing for Model Inference:} Data engineering enables the continuous capture and processing of data required for model inference. 
\end{enumerate}

\subsubsection{Definition of MLOps}
MLOps refers to a methodology aimed at managing the lifecycle of ML models, from design, training, and evaluation to distribution and deployment. It integrates DevOps, data engineering, and ML principles to ensure reliable and efficient deployment and maintenance of ML models in a controllable production setting. Essentially, MLOps = Data + DevOps + Model. MLOps applies DevOps practices to ML development and deployment, focusing on optimizing and accelerating processes. 
In summary, MLOps offers the following benefits for practical ML~\cite{kreuzberger2023machine}:

\begin{enumerate}
    \item \textit{Automation:} It automates the ML pipeline and CI/CD processes, reducing manual intervention, minimizing errors, and accelerating application delivery.
    \item \textit{Quality control:} It ensures consistent and reliable deployment and maintenance of ML models in a production environment, maintaining service quality standards and reducing the risk of errors.
    \item \textit{Unified workflow:} It provides a cohesive framework for managing the lifecycle of ML models, fostering collaboration and efficiency across teams.
\end{enumerate}

Currently, leading entities in the AI/ML domain, such as Microsoft, AWS, and Google, have widely adopted and implemented MLOps principles. Additionally, Databricks has recently introduced Unity Catalog. This supports the full lifecycle of an ML model by leveraging Unity Catalog’s capability to share assets across Databricks workspaces and trace lineage across both data and models~\cite{bigbookmlops}.

\subsection{The Evolutionary Trend of RAN}
This article focuses on MLOps implementation at the RAN level. RAN, the most critical component in wireless networks, includes the hardware and software that connects user equipment (UE) to the core network via radio links. It serves as the network entrance and provides services to UE. Recent advancements in RAN highlight the following important features:

\begin{enumerate}
    \item \textit{Virtualization} refers to the decoupling of hardware resources from the software functions in the RAN, abstracting RAN hardware and functionalities into virtualized environments. This allows for more flexible and efficient use of network resources.
    \item \textit{Disaggregation} is a recent trend involving separating network hardware and software components, allowing them to be sourced from different vendors or developed independently. A prime example is Open RAN, which splits the gNB into the radio unit (RU), distributed unit (DU) and central unit (CU).
    \item \textit{Hierarchical computing} involves the co-location of RAN and other computational entities, such as multi-access edge and fog computing, to optimize network performance, reduce latency, minimize transport overhead, and improve resource utilization in complicated networks.
    \item \textit{Intelligence} is a compelling ongoing research topic in network operations. As stated in~\cite{ituimt2030}, AI and communication are key pillars of the 6G scenarios. The adoption of AI/ML can fully harness the benefits of network virtualization, disaggregation, and hierarchical computing, thereby enhancing performance, improving energy efficiency, and fostering innovations. %in wireless communication systems.
    
\end{enumerate}

\subsection{Challenges of Applying MLOps in Advanced RAN}
\begin{figure}[t]
    \centering
    \includegraphics[width=0.9\linewidth]{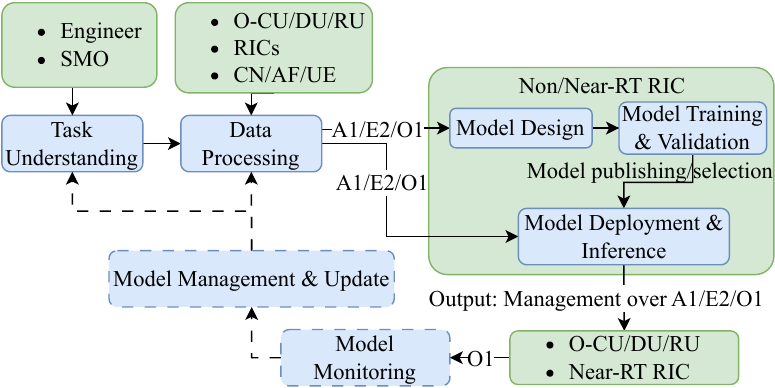}
    \vspace{-0.2cm}
    \caption{ML framework in Open RAN.}
    \vspace{-0.5cm}
    \label{fig:ML_in_Open_RAN}
\end{figure}

With the prevalence of AI/ML integration in the network, a holistic model's monitoring, management, and updating mechanism is foundational to support the correct and seamless functioning of the ML models in the dynamic environment over its lifecycle. The above can be delivered by an MLOps framework. However, MLOps performs well in a centralized cloud environment that is relatively simple and controllable, where the data originates from a singular source, and the models involved are straightforward.
Instead, applying MLOps to wireless systems becomes problematic as the scenarios get more complicated due to the following reasons:
\begin{enumerate}
    \item \textit{Data Variety:} 
    The data source is diversified. Data could come from the cloud, edge, RAN, application function (AF), and UEs. The hierarchical network architecture leads to a hard-to-define common data processing pipeline. Meanwhile, ML models are prone to being affected by input features, which are highly related to the model's executive environment under the wireless setting.
    \item \textit{Learning Variety:} 
    Network optimization may involve different learning paradigms, each requiring specific environmental settings. For instance, RL needs interaction with the network environment to explore the optimal policy, while FL must consider bias factors among clients, such as variations in communication channels, computational capability, and dataset balance.
    \item \textit{Deployment Variety:} 
    The deployment venues for AI/ML models are diverse, including the cloud, core network, RICs, and edge devices. Each venue requires a tailored AI/ML workflow, encompassing interface/API definitions and model distribution methods.
    \item \textit{Un-closed Ops Loop:} 
    The AI/ML training-deployment-update loop in wireless networks lacks a clear definition. Fig.~\ref{fig:ML_in_Open_RAN} illustrates examples of the ML framework in Open RAN. According to Open RAN specifications, tasks such as understanding, data processing, model training, and actuation can be carried out within its components. However, monitoring and management schemes are not standardized and are typically developed or implemented by the systems integrator.
\end{enumerate}

Therefore, general-purpose MLOps pipelines must be adjusted to accommodate the specific features of different learning paradigms. Moreover, the network should natively support customized MLOps mechanisms, considering effectiveness, timeliness, and sensitive data protection.

Embracing customized MLOps in advanced RAN offers clear benefits. It will streamline the development of automated and self-organized networks, unifying network control and management under a single pipeline. This integration enhances the scalability of AI/ML operations, fostering the development of AI-native networks and new communication paradigms.

In the following sections, we discuss the implementation of MLOps in RAN with different learning paradigms (RL, FL, and GenAI). We detail the key constraints of each learning approach and propose countermeasures to enhance the MLOps.

\section{MLOps for Reinforcement Learning: RLOps}
\label{sec:RLOps}
RL is a type of ML where an agent learns to make decisions by interacting with an environment. The agent receives feedback in the form of rewards based on its actions, and its goal is to learn the optimal actions that maximize cumulative reward over time. In deep RL (DRL), the deep neural network learns the optimal mapping between the state and action to maximize reward. 
In the context of RAN operation, DRL emerges as an appealing solution due to its inherent self-learning capabilities. DRL agents can learn the optimal RAN operational policy across various factors and time scales, facilitating the attainment of cross-layer/domain, long-term, and short-term balance optimization. Typical applications of DRL in RAN include wireless access control, traffic steering, baseband placement optimization, etc.

\subsection{Ops Challenges for RL}
Large-scale adoption of DRL in RAN is a challenge. That is because the policy learned by DRL is shaped by its interactive environment and optimization objectives, making the construction of the environment a critical factor influencing the learning process. This challenge is commonly known as the sim2real problem in DRL. Within the RAN system, sim2real pertains to system modeling issues, including factors such as reward delays, partial observations, and system variation/degradation over time. To ensure the successful deployment and operation of DRL models in practical RAN, it is important to guarantee the fidelity of the training environment and maintain it over time.
Conversely, DRL training is infamous for being time-consuming. Accelerating the DRL training process is essential to meeting diverse deployment needs across various scenarios.
These two key considerations raise challenges for the Ops of RL.

\subsection{RLOps}
\begin{figure}[t]
    \centering
    \includegraphics[width=1.0\linewidth]{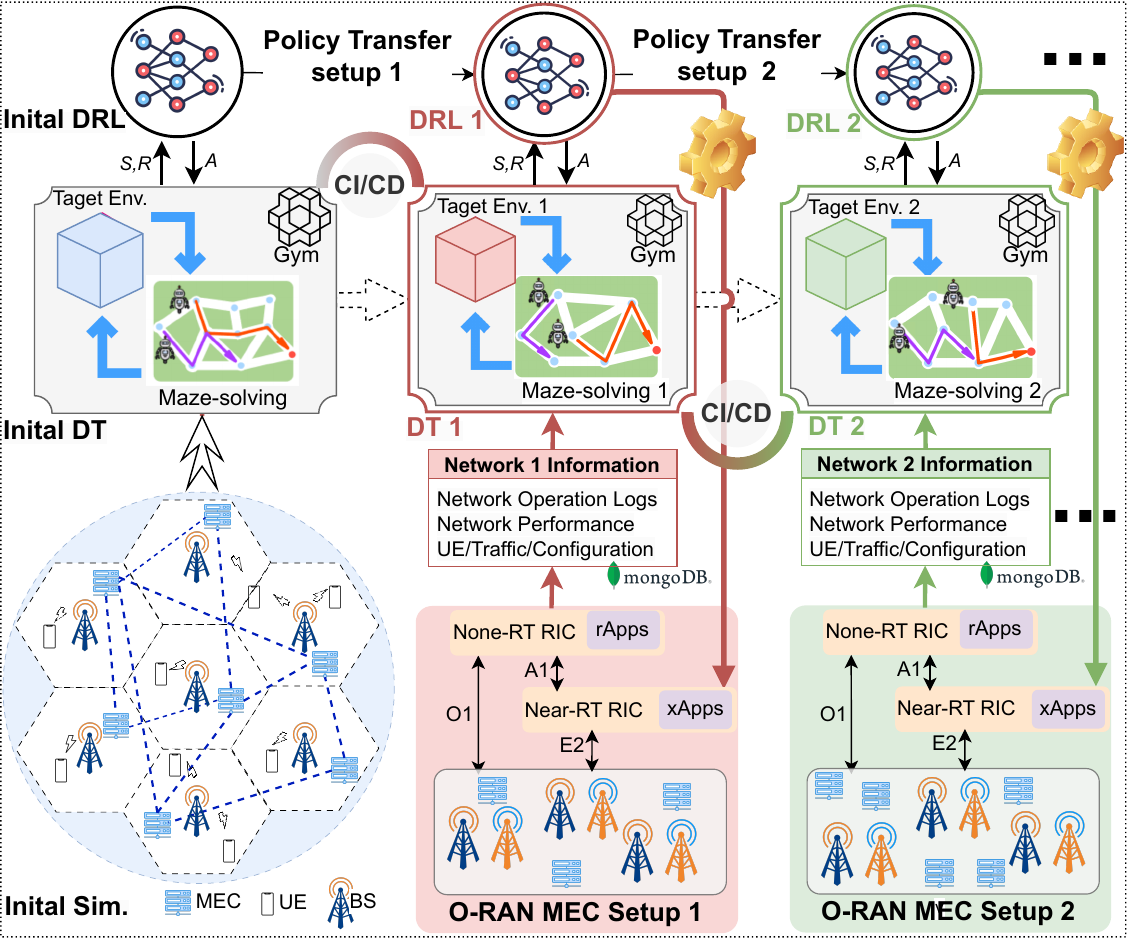}
    \vspace{-0.2cm}
    \caption{Illustration of an RLOps application for baseband placement across different Open RAN MEC setups.}
    \vspace{-0.2cm}
    \label{fig:RLOps}
\end{figure}
The concept of RLOps, introduced in the publication~\cite{li2022rlops}, presents a structured approach for RL models LCM, specifically tailored for intelligent Open RAN systems. The principles of RLOps outlined in their work offer a comprehensive framework, categorizing essential elements across design, development, operations, and safety/security considerations. This article aims to delve deeper into the analysis of RLOps by focusing on two key factors hindering the widespread adoption of DRL: environmental fidelity and training acceleration.
Integrating digital twins (DTs) within an RLOps framework is deemed a fundamental component in addressing such challenges. The DT concept and its practical applications have been extensively discussed as foundational techniques for developing 6G networks. The scope and functionality of digital modeling, APIs, and interface design have been explored in a wealth of literature. Essentially, a DT provides a high-fidelity virtual representation of a physical entity while facilitating communication with the real network setting for parameters, configurations, and pattern synchronization and updates. Thus, RLOps could leverage the DT as the environment for DRL training and updates.
Furthermore, tailored learning knowledge transfer techniques or modules should be incorporated into the RLOps pipeline to accelerate the iterative training or fine-tuning of DRL models. Representative schemes include transfer learning, incremental learning, knowledge distillation, domain adaptation, etc. The selection of specific techniques should be based on the characteristics of different DRL agents.

\subsection{RLOps in Open RAN: A Baseband Placement Case Study}

Placing baseband functions and user plane functions (UPF) in conjunction with multi-access edge computing (MEC) to accommodate diverse 5G services is a crucial challenge in modern network architectures. This usecase addresses the joint placement problem within the context of Open RAN, aiming to devise a function placement strategy that optimizes resource usage and reduces network power costs while maintaining service quality objectives.

Firstly, the function placement problem, including DU, CU-CP, CU-UP, and UPF, is modeled as a maze-solving task within a Markov decision process (MDP). A graph convolutional network (GCN) encoder is introduced, enabling the DRL agent to generalize across different network topologies. This approach unifies network features, reduces retraining needs, and enhances adaptability and scalability in diverse RAN architectures.
For the DRL training, a DT is built to serve as a customized setting for optimal RL policy exploration. This DT models \emph{1)} service types, \emph{2)} baseband function chains, and \emph{3)} MEC components along with their energy costs. More details of the problem sets can be found in~\cite{li2024netmind}.

In the pre-simulation stage, the DRL agent interacts with the DT to explore optimal actions, which include decisions on routing directions and function placements. The state in this DRL framework includes information on network resources, current placements, path lengths, and service requests, which vary from one deployment site to another. Therefore, in the deployment stage, this site-specific information must be updated in the DT accordingly to update the DRL agent, as shown in Fig.~\ref{fig:RLOps}. This DRL agent is deployed in the Near-RT RIC of Open RAN, utilising internal interfaces such as A1, O1, and E2 for messaging.

\begin{figure}[t]
    \centering
    \includegraphics[width=0.85\linewidth]{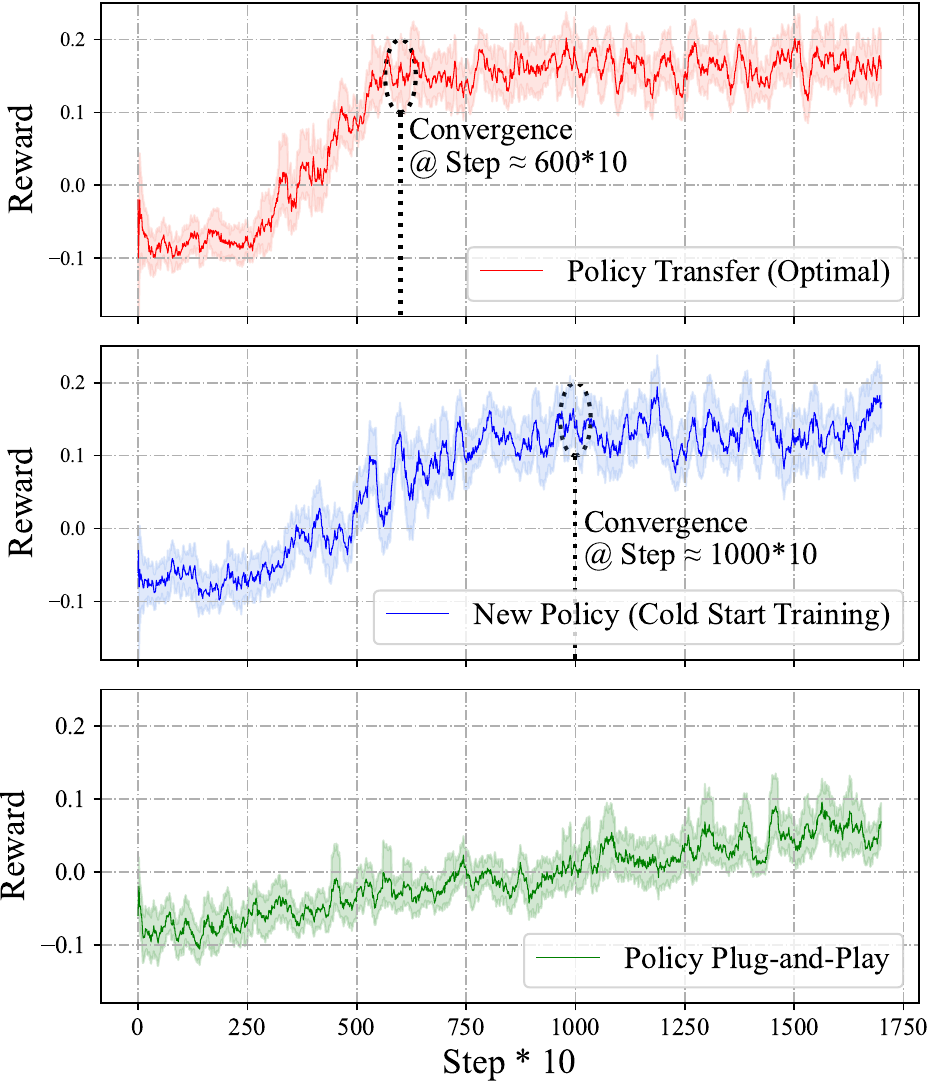}
    \vspace{-0.3cm}
    \caption{DRL Convergence Performance in New Open RAN MEC Setup with and without Policy Transfer.}
    \vspace{-0.5cm}
    \label{fig:result}
\end{figure}

Additionally, a DRL policy transfer technique, specifically incremental learning, is applied to speed up the fine-tuning process of the DRL agent from one deployment environment to another. As demonstrated in Fig.~\ref{fig:result}, when the policy is directly applied to a new Open RAN MEC setup, its performance degrades significantly (green line). However, when optimal policy transfer is applied, the new DRL convergence occurs around step 6000 (red line), which is a significant improvement compared to training a new agent from scratch, which converges at step 10000 (blue line).

\section{MLOps for Federated Learning: FedOps}
\label{sec:FedOps}
In FL, models are trained across multiple decentralized devices or servers. FL is commonly visible in wireless communication systems because it enables ML models to be trained directly on data at the edge without transmitting sensitive data to a central server. FL can reduce communication overhead, preserve user privacy, and allow for personalized model updates tailored to individual devices or network conditions. This paradigm can be flexibly applied to RAN settings with multiple computational layers, such as the edge, UE, and associated IoT devices. By applying FL in networks, distributed intelligence can be realized, and correspondingly, the wireless networks' efficiency and adaptability by enabling on-device model updating can be improved.

\subsection{Ops Challenges for FL}
The operational challenges of FL stem from inherent biases in wireless network systems. FL operates as a distributed training and centralized aggregation algorithm, where the effectiveness and convergence of global aggregation hinge on the characteristics of participating devices, representing the system's bias. This bias can be viewed from three perspectives: data, hardware, and communication link biases. Data bias pertains to the uneven distribution of data across clients, like non-independent and identical distribution (Non-IID). Hardware bias manifests as variations in hardware capabilities among devices, encompassing differences in local memory, processing speed, power constraints, etc. Communication link bias predominantly arises from the wireless channel, influenced by fluctuations in communication link quality and disparate bandwidth availability, which in turn causes asynchronous communication and affects the model parameter updating, aggregation, and broadcasting. The interplay of these biases necessitates delicate solutions in data engineering, hyperparameter configuration, model sharing, and synchronization. These aspects have traditionally been overlooked in standard MLOps practices, thus requiring FedOps to ensure the long-term reliability of FL models.

\subsection{FedOps}
The term of FedOps was introduced in~\cite{moon2024fedops}. It proposes a framework for analyzing client/system heterogeneity using a small subset of client data. Subsequently, appropriate clients are selected based on considerations of communication costs and overall model accuracy. While this work marks a promising beginning for exploring FedOps in engineering, it is important to address the long-term stability and robustness of FL operations beyond solely selecting clients.
In this article, we underscore the importance of implementing monitoring to mitigate hardware and communication biases and to enhance FedOps, particularly in wireless system environments.
While monitoring is a fundamental aspect of MLOps, its scope needs to be expanded for effective FedOps support. From the perspective of in-network FL model LCM, FL monitoring should encompass two key dimensions: \emph{1)} global-level monitoring and \emph{2)} client-level monitoring. Global monitoring in FedOps should be closely integrated with observable metrics of the wireless network, such as UE connectivity and channel quality. These metrics serve as crucial criteria for client selection and algorithm optimization. Client Monitoring in FedOps aims to ensure the proper functioning of clients. This involves examining the operational state of containers and clients themselves. Regular reports on corresponding metrics are then relayed to the global model.
The reporting and messaging mechanisms should leverage the internal data interfaces of wireless systems, such as A1, O1, and E2 in Open RAN, to reduce latency and improve efficiency.

\begin{figure*}[t]
    \centering
    \includegraphics[width=0.82\linewidth]{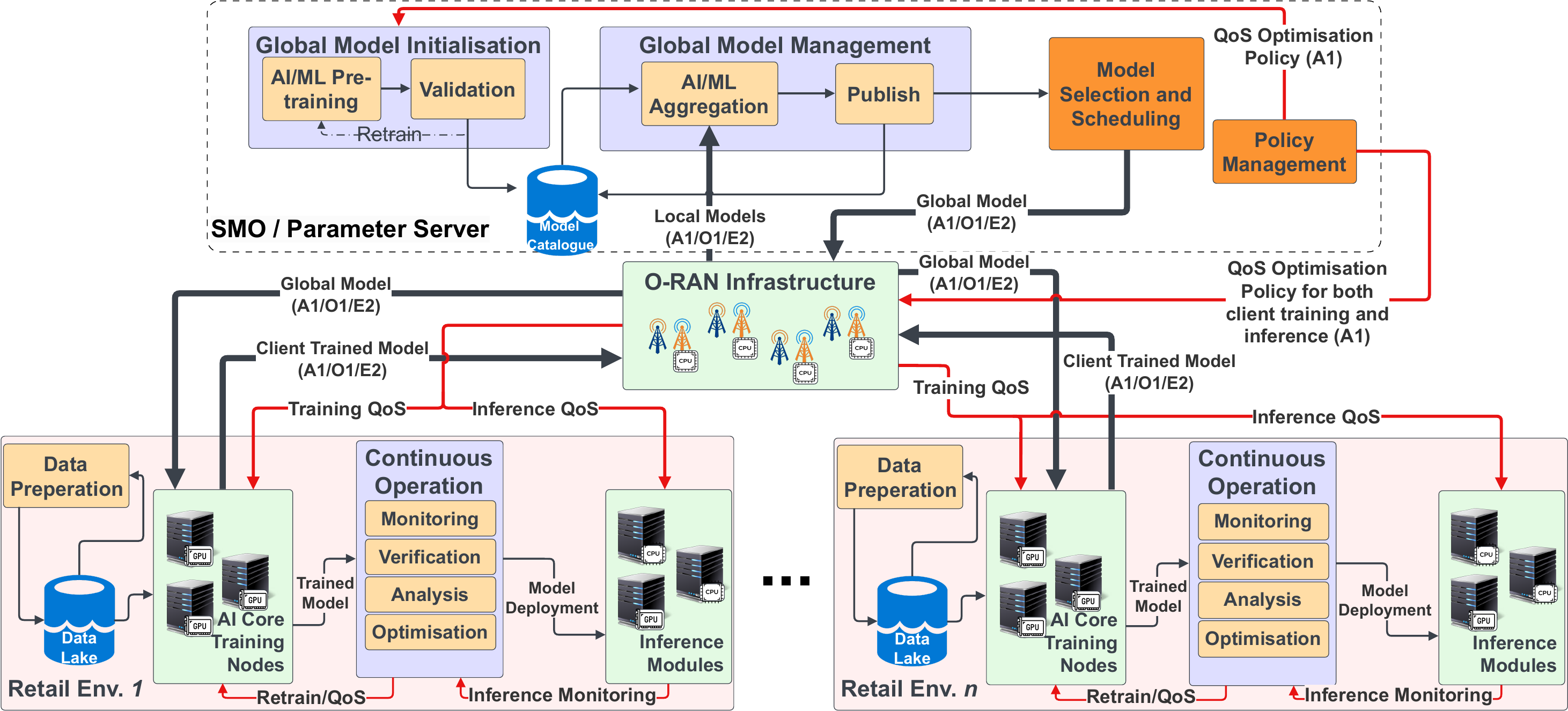}
    % \vspace{-0.7cm}
    \caption{Illustration of the Open RAN-based FLOps application in multiple retail environments with FL.}
    \vspace{-0.5cm}
    \label{fig:FLOps}
\end{figure*}
\subsection{FedOps in Retail: A Case Study}
FedOps enables robust and scalable deployment of FL methods agnostic to any particular domain. In this section, we consider a real-world use case for deploying FedOps in the context of dynamic and heterogeneous retail environments.
In the retail domain, two primary service categories emerge: \emph{1)} end-user services, encompassing personalized recommendation engines, dynamic pricing strategies, and trend identification algorithms, and \emph{2)} retail owner services, including critical applications like Point-of-Sale (PoS) device remote maintenance, theft prevention systems, and intelligent advertisement services.

Here we consider utilization of FedOps agnostic to any of these service categories. 
This could enable critical use cases for PoS devices, which handle sensitive financial transactions, have limited compute resources and are vulnerable to hardware failures, software glitches, and security breaches, necessitating robust monitoring and maintenance mechanisms. 
FedOps can enable the deployment of federated anomaly detection models that can detect anomalous behavior or performance issues in PoS devices without compromising customer data privacy.

The dataflow in this FedOps deployment begins with each PoS device collecting its operational data, such as transaction logs, system logs, and sensor readings etc.%, to the extent possible given its limited compute capabilities.
These can be stored on-premise for local, private processing (see Fig.~\ref{fig:FLOps}).
This locally stored data is then securely and privately used by the local edge device for further processing within the retail establishment, ensuring that sensitive customer data never leaves the premises, and preserving privacy.

At the local edge device, which serves as a \textit{client} in the FL system, the data from multiple \textit{endpoint} devices can be aggregated and used to train a local model, e.g., an anomaly detection model for PoS terminals maintenance.

Periodically, FedOps' client selection and scheduling module intelligently selects locally trained models from a subset of retail establishments (clients). This module employs strategies like clustering techniques or RL approaches to prioritize clients based on computational resources, network conditions, and data quality.
The global model is shared with the selected clients via the Open RAN infrastructure, as illustrated in Fig.~\ref{fig:FLOps}. The selected clients perform local training, ensuring QoS optimization for both training and inference policies. After local training, the updated client models are transmitted to the RIC over the same wireless link.

At the RIC/service management and orchestration (SMO) layer, FedOps orchestrates the secure aggregation of the locally trained models using efficient communication protocols tailored for FL, such as structured updates, quantization, and sparsification techniques. This aggregation process yields a globally trained model that captures collective knowledge from multiple retail establishments without exposing individual establishment or customer data.
The globally trained model is then distributed back to the participating retail establishments, where local edge devices can be used for inference. To further enhance privacy during the distribution phase, FedOps can employ differential privacy mechanisms, such as noise addition or data subsampling.

Throughout the entire dataflow, FedOps ensures efficient utilization of wireless resources by adapting communication strategies based on network conditions. It employs techniques like over-the-air aggregation protocols (e.g., coded computing, analog/hierarchical aggregation) to reduce communication overhead and minimize the impact of stragglers or disconnected clients through intelligent client selection algorithms.

By leveraging FedOps' federated anomaly detection capabilities and utilizing local edge devices as clients, retail establishments can proactively identify and address potential issues with PoS devices. This improves their reliability, security, and overall operational efficiency while maintaining strict privacy standards for customer data and accommodating the limited computing resources of individual PoS devices

\section{MLOps for Generative AI: GenOps}
\label{sec:GenOps}
GenAI involves techniques that generate new data instances, such as text, images, sounds, or videos, by leveraging patterns learned from existing data. Typical generative models include autoencoders, generative adversarial networks (GANs), and diffusion models, for modality generation. The remarkable success of large language models (LLMs) has sparked a surge in the development of generative models across various domains. Today, significant strides have been made in text-to-text, text-to-image/image-to-text, and even text-to-video generation.
This advancement has also drawn attention to its potential in wireless communication. The integration of GenAI into wireless networks, for its design, configuration, and operation, has become a focal point of research~\cite{bariah2024large}. One prominent approach involves goal-oriented and semantic communication, where information is encoded into a latent space by an encoder and then reconstructed by a decoder. Any of the aforementioned generative models can serve as the encoder-decoder pair~\cite{10453123}.
By incorporating GenAI in different network layers, networks can enhance operational efficiency while reducing overhead and energy consumption.

\subsection{Ops Challenges for GenAI}
The seamless integration of GenAI and wireless networks at the protocol level is still being explored. 
Here, we examine the potential challenges of GenAI in-network operation and LCM from the following perspectives:
\begin{itemize}
    \item \textit{Vast knowledge base (KB)}: The KB comprises essential data and features for generating coherent and relevant content. Its context is crucial for producing accurate, contextually appropriate responses and interactions from the given input. Compared to their larger cloud-based counterparts, optimally usable GenAIs in vertical applications are task-specific and trained with extensive knowledge bases. This poses significant cost and energy challenges for GenOps.
    \item \textit{Extracting/Incorporating human value information:} A notable feature of large generative models is their ability to gather human feedback to enhance performance and align with human preferences and values. Techniques such as human-in-the-loop (HITL) and reinforcement learning with human feedback (RLHF) enable continuous improvement toward relevant and coherent responses. However, integrating these models into networks complicates scenarios, as user feedback collection must pass through the network. Extracting, interpreting, and incorporating human value in a network poses a long-term operational challenge for CI/CD.
    \item \textit{Model integration:} Integrating GenAIs into networks presents challenges due to their size and costs. Additionally, combining various purpose-specific generative models raises concerns about both performance guarantees, potentially compromising the superiority of single task-specific GenAI models, and energy/resource consumption, impacting sustainability efforts. 
    
\end{itemize}

\subsection{GenOps for 6G}
In this article, GenOps refers to a pipeline aimed at seamlessly integrating GenAI applications into future 6G networks, ensuring long-term reliable operation through customization and maintenance of GenAI models.
As a pioneering effort, we summarize key solutions to the aforementioned challenges, differentiating GenOps from other Ops.
Firstly, GenOps must facilitate a loop for updating the knowledge base and fine-tuning generative models. This iterative process occurs external to the network and is initiated either by the network's internal task demands or external safety/security requirements. Secondly, GenOps should integrate human feedback collection over networks to facilitate models' self-adaption. This can be facilitated by, e.g., web-based feedback forms or RESTful APIs for mobile apps. Lastly, GenOps should be capable of offering different strategies (trade-off considerations) based on the scale of the generative model. For smaller models, containerized onboarding is feasible, whereas for larger models, alternative solutions such as integrating them through a subscription-based model should be considered.

Generative models can also be monetized by exposing them through APIs from the marketplace, which can enable tight integration with networks and different business models based on requested frequency and service level. Looking ahead, integrating these models with intent-based networking open service platforms could further enable a "network of networks" as a service paradigm. This approach would facilitate sophisticated orchestration of services across various networks operated by different entities, employing diverse technologies like satellite, terrestrial, and high-altitude platforms in a seamlessly integrated manner.

\section{Conclusions}
\label{sec:conclusions}
MLOps within the RAN represent an exciting and dynamic field of research, yet standardized practices remain nascent. While 3GPP has initiated discussions on a generic mechanism for overseeing ML training, particularly in contexts of 5G Core and RAN, detailed exploration of diverse AI/ML learning paradigms and their vertical applications remains a gap.
This article introduces MLOps principles tailored specifically for the RAN, encompassing diverse learning paradigms through RLOps, FedOps, and GenOps. We outline associated challenges with each approach and propose potential solutions.
The authors anticipate that the outlined Ops principles will contribute to the future development of a standardized in-network MLOps framework. This framework is essential for advancing the reliability and efficacy of AI/ML applications within the RAN and beyond.

\bibliographystyle{IEEEtran} %
%\balance
%\bibliographystyle{plainnat} 
\bibliography{IEEEabrv,references} 

\begin{thebibliography}{10}
\providecommand{\url}[1]{#1}
\csname url@samestyle\endcsname
\providecommand{\newblock}{\relax}
\providecommand{\bibinfo}[2]{#2}
\providecommand{\BIBentrySTDinterwordspacing}{\spaceskip=0pt\relax}
\providecommand{\BIBentryALTinterwordstretchfactor}{4}
\providecommand{\BIBentryALTinterwordspacing}{\spaceskip=\fontdimen2\font plus
\BIBentryALTinterwordstretchfactor\fontdimen3\font minus \fontdimen4\font\relax}
\providecommand{\BIBforeignlanguage}[2]{{%
\expandafter\ifx\csname l@#1\endcsname\relax
\typeout{** WARNING: IEEEtran.bst: No hyphenation pattern has been}%
\typeout{** loaded for the language `#1'. Using the pattern for}%
\typeout{** the default language instead.}%
\else
\language=\csname l@#1\endcsname
\fi
#2}}
\providecommand{\BIBdecl}{\relax}
\BIBdecl

\bibitem{letaief2019roadmap}
K.~B. Letaief \emph{et~al.}, ``{The Roadmap to 6G: AI Empowered Wireless Networks},'' \emph{IEEE Commun. Mag.}, vol.~57, no.~8, pp. 84--90, 2019.

\bibitem{ituimt2030}
``{IMT-2030 Vision - International Telecommunication Union (ITU)},'' \url{https://www.itu.int/en/ITU-R/study-groups/rsg5/rwp5d/imt-2030/Pages/default.aspx}, accessed: July 02, 2024.

\bibitem{rezazadeh2024sliceops}
F.~Rezazadeh \emph{et~al.}, ``{SliceOps: Explainable MLOps for Streamlined Automation-Native 6G Networks},'' \emph{IEEE Wirel. Commun.}, 2024.

\bibitem{3gpp23.700}
{3GPP}, ``{Study on 5G system support for AI/ML-based services},'' {3rd Generation Partnership Project (3GPP)}, {Technical Report (TR)} 23.700-80, Release 18.

\bibitem{3gpp28.105}
------, ``{{Management and orchestration; Artificial Intelligence/ Machine Learning (AI/ML) management}},'' {3rd Generation Partnership Project (3GPP)}, {Technical Specification (TS)} 28.105, Release 17.

\bibitem{3gpp_website}
\BIBentryALTinterwordspacing
------. {3GPP Technologies - AI/ML Management for 5G Systems}. Accessed: July 02, 2024. [Online]. Available: \url{https://www.3gpp.org/technologies/ai-ml-management}
\BIBentrySTDinterwordspacing

\bibitem{polese2023understanding}
M.~Polese \emph{et~al.}, ``{Understanding O-RAN: Architecture, Interfaces, Algorithms, Security, and Research Challenges},'' \emph{IEEE Commun. Surv. Tutor.}, vol.~25, no.~2, pp. 1376--1411, 2023.

\bibitem{strinati20216g}
E.~C. Strinati and S.~Barbarossa, ``{6G networks: Beyond Shannon towards semantic and goal-oriented communications},'' \emph{Computer Networks}, vol. 190, p. 107930, 2021.

\bibitem{kreuzberger2023machine}
D.~Kreuzberger, N.~K{\"u}hl, and S.~Hirschl, ``{Machine Learning Operations (MLOps): Overview, Definition, and Architecture},'' \emph{IEEE Access}, 2023.

\bibitem{bigbookmlops}
J.~Bradley, R.~Kurlansik, M.~Thomson, and N.~Turbitt, \emph{The Big Book of MLOps eBook - 2nd edition}.\hskip 1em plus 0.5em minus 0.4em\relax Databricks, 2023.

\bibitem{li2022rlops}
P.~Li \emph{et~al.}, ``{RLops: Development Life-cycle of Reinforcement Learning Aided Open RAN},'' \emph{IEEE Access}, vol.~10, pp. 113\,808--113\,826, 2022.

\bibitem{li2024netmind}
H.~Li \emph{et~al.}, ``{NetMind: Adaptive RAN Baseband Function Placement by GCN Encoding and Maze-solving DRL},'' \emph{arXiv preprint arXiv:2401.06722}, 2024.

\bibitem{moon2024fedops}
J.~Moon, S.~Yang, and K.~Lee, ``{FedOps: A Platform of Federated Learning Operations with Heterogeneity Management},'' \emph{IEEE Access}, 2024.

\bibitem{bariah2024large}
L.~Bariah \emph{et~al.}, ``{Large Generative AI Models for Telecom: The Next Big Thing?}'' \emph{IEEE Commun. Mag.}, 2024.

\bibitem{10453123}
P.~Li and A.~Aijaz, ``{Open RAN meets Semantic Communications: A Synergistic Match for Open, Intelligent, and Knowledge-Driven 6G},'' in \emph{Proc. of IEEE CSCN}, 2023, pp. 87--93.

\end{thebibliography}
\vspace{0.5cm}
\small{\noindent \textbf{Peizheng Li} is a Research Engineer at the Bristol Research and Innovation Laboratory, Toshiba Research Europe Ltd.}

\small{\noindent \textbf{Ioannis Mavromatis} is a Lead 5G/Future Networks Technologist at Digital Catapult.}

\small{\noindent \textbf{Tim Farnham} is Chief Research Fellow with the Bristol Research and Innovation Laboratory, Toshiba Europe Ltd. His recent research includes applying digital twins and data spaces for optimizing wireless, water distribution and energy network systems. Collaborative evaluation of these techniques have been performed within EU Horizon and UK projects.}

\small{\noindent \textbf{Adnan Aijaz} is currently the Programme
Leader for Beyond 5G at the Bristol Research and Innovation Laboratory,
Toshiba Research Europe Ltd., U.K. His recent research interests include
5G/6G systems, Open RAN, time-sensitive networking, high-altitude
platforms, and robotics and autonomous systems.}

\small{\noindent \textbf{Aftab Khan} is currently the Distributed AI Programme Leader with Bristol Research and Innovation Laboratory, Toshiba Europe Ltd., U.K. His research interests include federated learning, AI-driven cyber security, computational behavior analysis, and pattern recognition.}

\end{document}